\documentclass[%
 preprint,%
 amssymb, amsmath,%
 aip,jcp,%cha,%
]{revtex4-1}

\usepackage{graphicx}% Include figure files
\usepackage{dcolumn}% Align table columns on decimal point
\usepackage{bm}% bold math
\usepackage[table,xcdraw]{xcolor}
\usepackage[utf8]{inputenc}
\usepackage[T1]{fontenc}
\usepackage{mathptmx}
\usepackage{etoolbox}

\usepackage{color, colortbl}
\makeatletter
\def\@email#1#2{%
 \endgroup
 \patchcmd{\titleblock@produce}
  {\frontmatter@RRAPformat}
  {\frontmatter@RRAPformat{\produce@RRAP{*#1\href{mailto:#2}{#2}}}\frontmatter@RRAPformat}
  {}{}
}%
\makeatother
\begin{document}
\newcommand{\beginsupplement}{%
        \setcounter{table}{0}
        \renewcommand{\thetable}{S\arabic{table}}%
        \setcounter{figure}{0}
        \renewcommand{\thefigure}{S\arabic{figure}}%
     }

\title{Non adiabatic dynamics of photoexcited cyclobutanone: predicting structural measurements from trajectory surface hopping with XMS-CASPT2 simulations}
% Force line breaks with \\
\author{Patricia Vindel-Zandbergen}%
 \email{patricia.vindel@nyu.edu}
\affiliation{Department of
Chemistry, New York University, New York, New York 10003, USA 
%\\This line break forced with \textbackslash\textbackslash
}%

\author{Jesús González-Vázquez}

 \affiliation{Departamento de Química,
Universidad Autónoma de Madrid, Cantoblanco 28049 Madrid (Spain)}%Lines break automatically or can be forced with \\


\begin{abstract}
For years, theoretical calculations and scalable computer simulations  have complemented ultrafast experiments as they offer the advantage to overcome experimental restrictions and have access to the whole dynamics. This synergy between theory and experiment promises to yield a deeper understanding of photochemical processes, offering valuable insights into the behavior of complex systems at the molecular level. However, the capacity of theoretical models to predict ultrafast experimental outcomes has remained largely unexplored. In this work, we aim to predict the electron diffraction signals of an upcoming ultrafast photochemical experiment using high-level electronic structure calculations and non adiabatic dynamics simulations. In particular, we perform trajectory surface hopping with XMS-CASPT2 simulations for understanding the photodissociation of cyclobutanone upon excitation at 200 nm. Spin-orbit couplings are considered for investigating the role of triplet states. Our simulations capture the bond cleavage after ultrafast relaxation from the
3s-Rydberg state leading to the formation of the previously
observed primary photoproducts: CO + cyclopropane/propene (\textbf{C3} products), ketene and ethene (\textbf{C2} products). The ratio of the \textbf{C3:C2} products
is found to be about 1.2. Within 800 fs, the majority of trajectories transition to their electronic ground state,  with a small fraction conserving the initial cyclobutanone ring structure. We found a minimal influence of triplet states during the early stages of the dynamics, with their significance increasing at later times. We simulate MeV-ultrafast electron diffraction (UED) patterns from our trajectory results, linking the observed features with specific photoproducts and the underlying structural dynamics. Our analysis reveals highly intense features in the UED signals corresponding to the photochemical processes of cyclobutanone. These features offer valuable insights into the experimental monitoring of ring opening dynamics and the formation of \textbf{C3} and \textbf{C2} photoproducts.  
\end{abstract}

\maketitle

\section{\label{sec:Intro}Introduction}
Understanding the structure and dynamics of chemical processes at the molecular level is a key step toward the
design of materials with desired properties or the efficient control of chemical reactions.
Recent progress in laser technology and imaging techniques have revolutionized the study of ultrafast dynamics by providing incredibly short and intense X-ray pulses and the possibility to probe structural changes in molecules after photoexcitation with sub-Angstrom
spatial and few-fs temporal resolution\cite{xfel_2021}. 
Direct imaging of
the motion of individual atoms in photoinduced reactions still remains challenging and often relies on the comparison with high-level theoretical calculations\cite{PRX_Weinacht, Borne_2023,Chakraborty_20, boll_x-ray_2022,Wolf_23, Basile_24,JGV_24}.
Ab initio non adiabatic molecular dynamics (MD) simulation methods offer a universal tool to model ultrafast photochemical processes. 
Some examples include photosynthesis~\cite{TMF11}, radiation damage of DNA under UV light~\cite{Satzger10196,B815602F,RMGSG12} and charge dynamics in solar cell materials~\cite{AP14,CZNTP17}. The simulation of non adiabatic dynamics and its connection to experimental data pose significant challenges due to various factors such as the large number of degrees of freedom, the need for expensive and accurate ab initio calculations, consideration of large time scales, and the effects of the environment or nuclear quantum effects. Theoretical methods used for this purpose are inherently approximate and often involve considerable numerical expense, particularly when applied to realistic molecular systems with tens to hundreds of atoms, and to simulate dynamical time scales ranging from tens of femtoseconds to picoseconds.
Choosing an appropriate method becomes a trade-off between accuracy and computational cost, with uncertainties in the individual contributions to the total error bar often poorly defined or unknown. For instance, in describing nonradiative relaxation time scales, it is often unclear whether inaccuracies arise from electronic structure methodology, resulting in incorrect gaps between electronic states, or from the non adiabatic molecular dynamics method itself, which may evaluate incorrect transfer rates due to the common issue of missing coherence phenomena\cite{GP07}.
As a result, non adiabatic molecular dynamics simulations typically provide qualitative insights, and verifying computational results against spectroscopic data becomes highly desirable whenever feasible. 

A recent question that has arisen is whether non adiabatic dynamics simulations can definitively predict experimental outcomes.
This study aims to address this question by assessing whether simulations of excited state dynamics can accurately predict the results of an upcoming ultrafast pump-probe gas-phase experiment before it is conducted. In this experiment, cyclobutanone will be excited with 200 nm light, and electron diffraction images will be captured at various time delays, providing direct structural information as a function of time. 

Cyclobutanone (CB) has been the subject of numerous experimental and theoretical studies because of  its unusual photochemical
behavior, due to the strain effect from its four-membered
ring (see Ref.\citenum{liu_new_2016} and references therein). 
Two electronic transitions have been identified. The S$_0\rightarrow$S$_1$ band, with its absorption maximum around 280 nm, attributed to the symmetry-forbidden $n\rightarrow\pi^*$ transition \cite{hemminger_laser-excited_1973}, and the S$_0\rightarrow$S$_2$ band, centered at approximately 195 nm, assigned to the $n\rightarrow3s$ Rydberg transition \cite{otoole_vacuum-ultraviolet_1991}.
The photodynamics following excitation to its first excited state has been extensively investigated, both experimentally and theoretically \cite{xia_excited-state_2015,liu_new_2016,diau_femtochemistry_2001,hemminger_laser-excited_1973}. Moreover, dynamics upon excitation to the 3s-Rydberg state have also been explored \cite{kuhlman_symmetry_2012,,diau_femtochemistry_2001}. However, a comprehensive description of the photochemistry of cyclobutanone (CB), including the relaxation dynamics and the involved mechanisms such as the influence of triplet states, has not been previously been performed for this particular process.
In this article, as we await the results of the experiment, we offer a thorough examination of the ultrafast photodissociation dynamics of CB. We employ trajectory surface hopping simulations with XMS-CASPT2 and include spin-orbit couplings. Furthermore, we simulate ultrafast electron diffraction patterns for comparison with forthcoming experiments.

 This work is organized as follows: Sec. \ref{sec:Methods} outlines the theoretical methodology, computational details, and experimental setup.
 In Sec. \ref{sec:results} we present and discuss the numerical findings derived from the electronic structure and trajectory surface hopping simulations of cyclobutanone, along with the experimental observables derived from these calculations. Finally, in Sec. \ref{sec:conc} we summarize the key findings and insights extracted from this work.

\section{\label{sec:Methods}Computational details}
\subsection{\label{sec:ElectronicStruc}Electronic structure calculations}
The ground state equilibrium structure and vertical excitation energies of cyclobutanone have been previously determined at different levels of theory\cite{kuhlman_symmetry_2012,xia_excited-state_2015}. In our study,
we employed Møller-Plesset second-order perturbation theory (MP2)\cite{mp2} 
with the aug-cc-pvtz basis set\cite{augpvdz}. 
%This optimization was done using the BAGEL\cite{BAGEl, BAGEL_2} code. 
%%% double z or triple z... not Molcas?
The frequencies and normal modes were calculated at the same  level of theory. 
For the optimized geometry, we computed 
%several electronic states, 
vertical excitation energies and oscillator strengths using multireference methods based on state-averaged complete active space self-consistent field (SA-CASSCF) wavefunctions\cite{CASSCF} and benchmarked them against  experimental data\cite{otoole_vacuum-ultraviolet_1991,diau_femtochemistry_2001,hemminger_laser-excited_1973}.

The selection of the orbitals included in the CASSCF active space is chosen for the correct description 
of the excited states of CB, in particular, the $n\rightarrow 3s$ transition, and the possible fragmentation
processes. 
\begin{table}[h!]
\resizebox{\columnwidth}{!}{%
\begin{tabular}{cccccccc}
\hline\hline
\rowcolor[HTML]{FFFFFF} 
{\textbf{State}} & {\textbf{Character}} & { \textbf{Experiment}} & \multicolumn{3}{c}{{\textbf{SA(4)-CAS(4,4)}}} & { \textbf{SA(4)-CAS(8,8)}} & { \textbf{SA(4)-CAS(12-12)}} \\ \hline\hline
\textbf{}                             & \textbf{}                                 & \textbf{}                                  & \textbf{aug-cc-pvdz}         & \textbf{aug-cc-pvtz}         & \textbf{ano-rcc-pvdz}        & \textbf{aug-cc-pvdz}                           & \textbf{aug-cc-pvdz}                             \\ \hline
\textbf{S$_1$}                           & $n\rightarrow\pi^*$                                    & 4.4                                        & 4.34 (0.000)                         & 4.34                         & 4.40                         & 4.34 (0.000)                                           & 4.23 (0.000)                                             \\
\textbf{S$_2$}                           & $n\rightarrow3s$                                       & 6.3                                        & 6.32 (0.011)                        & 6.28                         & 7.03                        & 6.32 (0.013)                                           & 6.45 (0.015)                                            \\
\textbf{S$_3$}                           & $\pi\rightarrow3s$                                 &                                            & 8.84 (0.026)                        & 8.79                         & 9.52                         & 8.84 (0.026)                                            & 9.57 (4.4 E-5)                                             \\
\textbf{T$_1$}                           &    $n\rightarrow\pi^*$                                      &                                            & 3.90                        & 3.91                         & 3.97                         & 3.90                                           & 4.01                                             \\
\textbf{T$_2$}                           & $\pi\rightarrow\pi^*$                                         &                                            & 5.42                         & 5.40                         & 5.49                         & 5.42                                           & 5.77                                             \\
\textbf{T$_3$}                           &   $n\rightarrow3s$                                        &                                            & 6.23                         & 6.20                         & 6.91                         & 6.23                                           & 6.37                                             \\
\textbf{T$_4$}                           &     $\pi\rightarrow3s$                                      &                                            & 8.80                         & 8.79                         & 9.52                         & 8.80                                           & 9.31                                             \\ 
                                      &                                           &                                            &                              &                              &                              &                                                &                                                  \\ \hline\hline
\textbf{State}                        & \textbf{Character}                        & \textbf{Experiment}                        & \multicolumn{3}{c}{\textbf{XMS-CASPT2 (4,4)}}                                              & \textbf{XMS-CASPT2 (8,8)}                      & \textbf{XMS-CASPT2 (12,12)}                      \\ \hline\hline
\textbf{}                             & \textbf{}                                 & \textbf{}                                  & \textbf{aug-cc-pvdz}         & \textbf{aug-cc-pvtz}         & \textbf{ano-rcc-pvdz}        & \textbf{aug-cc-pvdz}                           & \textbf{aug-cc-pvdz}                             \\ \hline
\textbf{S$_1$}                           & $n\rightarrow\pi^*$                                    & 4.4                                        & 4.13 (0.000)                         & 4.08                         & 4.17                         & 4.17 (0.000)                                           & 4.15 (0.000)                                             \\
\textbf{S$_2$}                           & $n\rightarrow3s$                                   & 6.3                                        & 6.48 (0.011)                         & 6.73                         & 7.01                         & 6.36 (0.013)                                            & 6.13 (0.014)                                            \\
\textbf{S$_3$}                           & $\pi\rightarrow3s$                                 &                                            & 9.67 (0.031)                         & 9.85                         & 10.14                        & 9.85 (0.024)                                           & 8.65 (4.5E-5)                                            \\
\textbf{T$_1$}                           &    $n\rightarrow\pi^*$                                       &                                            & 3.78                         & 3.76                         & 3.82                         & 3.87                                           & 3.81                                             \\
\textbf{T$_2$}                           &  $\pi\rightarrow\pi^*$                                         &                                            & 6.18                         & 6.14                         & 6.26                         & 6.18                                           & 5.97                                             \\
\textbf{T$_3$}                           &   $n\rightarrow3s$                                        &                                            & 6.31                         & 6.55                         & 6.83                         & 6.25                                           & 6.15                                             \\
\textbf{T$_4$}                           &  $\pi\rightarrow3s$                                         &                                            & 9.69                         & 9.86                         & 10.28                        & 9.78                                           & 8.38                                            
\end{tabular}%
}
\caption{Vertical excitation energies (eV) and oscillator strengths, shown in parenthesis, of CB at the equilibrium geometry computed for different active spaces and basis sets at the SA(4/4)-CASSCF/XMS-CASPT2 levels of theory and experimental absorption maxima values taken from Refs.\citenum{otoole_vacuum-ultraviolet_1991,diau_femtochemistry_2001,hemminger_laser-excited_1973}. The CASSCF active spaces are denoted in the format $(e,o)$ where $e$ and $o$ represent
the number of electrons and orbitals in the active space, respectively.}
\label{table:elec_struc}
\end{table}
We performed several SA-CASSCF calculations using different sets of active spaces and basis sets to determine the optimal level of theory for the non adiabatic dynamics simulations. 
The following active spaces were tested: (1) four electrons in four orbitals, denoted as CAS(4,4), which includes the $\pi$ and $\pi*$ orbitals of the carbonyl group, the non-bonding orbital at the oxygen atom and the 3s Rydberg orbital; (2) eight electrons in eight orbitals, CAS(8,8), in which we added the $\sigma$ and $\sigma*$ orbitals of the C-C bonds between the carbonyl carbon and the $\alpha$-carbons; (3) twelve electrons in twelve orbitals, CAS(12,12), also including the 4 $\sigma$ and $\sigma*$ orbitals of the C-C bonds between the $\alpha$-carbons and the $\beta$-carbons to the former. 
The orbitals are shown in Figs. \ref{Fig:orbitals_CB_12} and \ref{Fig:orbitals_CB_8} of supplementary information.
We averaged over 4 states of each multiplicity -four singlets and four triplets- to account for the $n\rightarrow\pi^*$ and $n\rightarrow 3s$ transitions and  possible contributions of the $\pi\rightarrow 3s$ transitions in the dynamics when the CO photoproduct is formed.   
To account for dynamic correlation, the energies were corrected by extended multi-state complete active space with second order perturbation (XMS-CASPT2) theory with an imaginary shift of 0.3 and IPEA shift of zero\cite{SGCW11_XMSCASPT2}. 
\begin{figure}
\includegraphics[width=0.6\textwidth]{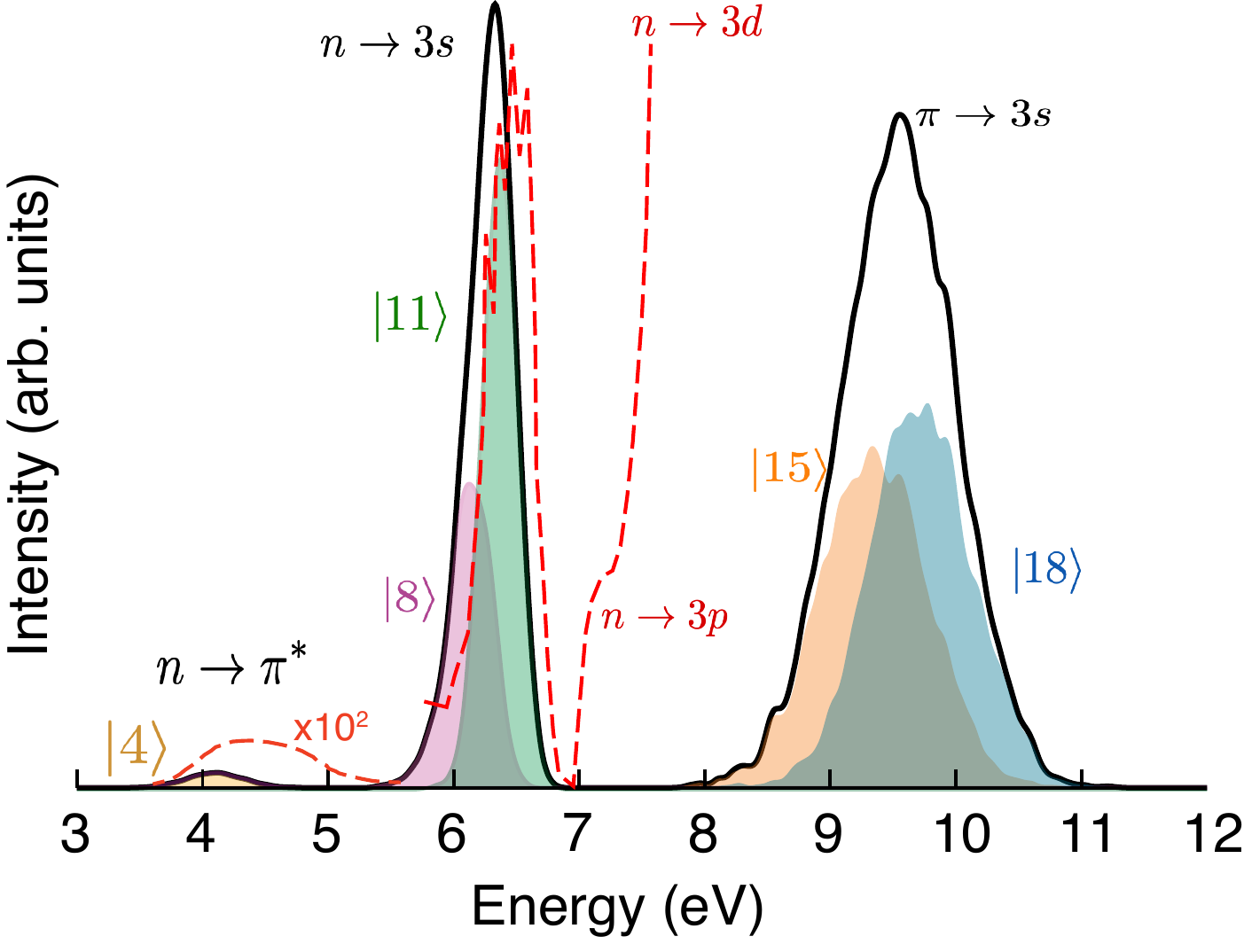}
\caption{Experimental (red dashed lines) and theoretical (black solid lines) absorption spectra computed at SA(4)-XMS-CASPT2(8,8)/aug-cc-pvdz level including SOCs, from  excitation of 10000 initial conditions.  The shaded curves represent the contributions of each spin-adiabatic state to every theoretical absorption peak. The absence of intensity in the $(n,3p)$ and $(n,3d)$ regions in the computed spectrum stems from our omission of the 3p and 3d states in our calculations. The absorption peak between 8 to 11 eV corresponds to a $\pi\rightarrow 3s$ transition, situated at energies far beyond the experimental excitation window of interest (200 nm). The experimental absorption spectrum has been from Refs. \cite{hemminger_laser-excited_1973,diau_femtochemistry_2001,otoole_vacuum-ultraviolet_1991} and the intensities in the region from peak at 3.7 to 5.5 eV have been multiplied by a factor of 100 for a better visualization. }
\label{fig:abs_spectra}
\end{figure}
For a correct description of the diffuse Rydberg states, we tested the accuracy of our calculations using aug-cc-pvdz and aug-cc-pvtz basis sets in the SA(4)-CAS(4,4) calculations. 
Since we include singlet-triplet transitions in our dynamics calculations, we consider relativistic effects  by including 
the spin-orbit couplings (SOCs) with the restrictive
active space interaction RASSI\cite{RASSI_CPL_02} and atomic
mean field integral AMFI\cite{AMFI} formalisms.  
We performed additional calculations using ano-rcc\cite{ANO_RCC_1,ANO_RCC_2} contracted to a doble zeta polarized (ano-rcc-vdzp), to account for relativistic core contributions.
While the energy of the valence states was very similar on both cases, the Rydberg excitation is overestimated on the ano-rcc-vdzp basis set, probably due to the lack of diffuse basis function on the basis set.

The CASSCF and XMS-CASPT2 calculations were performed
using the OpenMolcas code\cite{openmolcas23}.
Vertical excitation energies and oscillator strengths at different levels of theory are listed in Table \ref{table:elec_struc} and compared with experimentally available data\cite{hemminger_laser-excited_1973,otoole_vacuum-ultraviolet_1991,drurylessard_ring_1978}. 
%For simplicity, we include only XMS-CASPT2 energies for CAS(8,8) and CAS(12,12) active spaces, averaging over 4 singlets and 4 triplets, labeled as SA(4/4) at the optimized geometry and for a aug-cc-pvdz basis set, and compare with experimental available data\cite{hemminger_laser-excited_1973,otoole_vacuum-ultraviolet_1991,drury-lessard_ring_1978}. Additional information regarding the energies computed at different levels of theory can be found in Table \ref{Table_SI_methods}. 

Concerning computational efficiency and accuracy, the dynamics were finally performed at XMS-CASPT2/aug-cc-pvdz level with SA(4/4)-CAS(8,8) and including SOCs. 
The selected method strikes a balance, allowing us to accurately describe the excited states of interest and the $\alpha$-cleavage dynamics while ensuring that our calculations remain feasible within a reasonable time frame.

\subsection{\label{sec:TSH}Trajectory surface hopping dynamics}
We simulate the photodynamics of CB upon excitation at $200$ nm using trajectory surface hopping\cite{T90,SJLP16,Prezhdo_JPCL2016}. This methodology has a favorable balance between computational efficiency and accuracy due to its semiclassical character. 
It can be combined with on-the-fly \textit{ab initio} electronic structure calculations, yielding results -electronic populations, energies, nuclear positions and velocities- that readily correlate with experimental observations. 
In this approach, 
%the time evolution of the system is followed using a mix of quantum and classical dynamics, where 
the nuclei are treated classically while keeping the quantum nature of the electrons. Essentially, an ensemble of classical nuclear trajectories is evolved on adiabatic electronic potential energy surfaces (PES), where the surfaces represent different electronic states of the system. 
The nuclei follow Newton's equations of motion and the system can switch between surfaces stochastically during the propagation, accounting for transitions between electronic states induced by non adiabaticity (including spin-orbit coupling). 
The electronic wavefunction is propagated by the time-dependent Schrödinger equation.
For a detailed description of the method, recent progress and applications, the reader is referred to Refs. \citenum{MAI_SH, CB18,Tretiak_SH, Linjun_SH}. 

We simulate the non adiabatic excited state dynamics using the trajectory surface hopping (TSH) as implemented in the SHARC method (suface-hopping including arbitrary couplings) \cite{RMGSG_11_SHARC}, where the Hamiltonian and spin-free gradients were described using the electronic structure methodology previously described.  
We employed a
velocity-Verlet algorithm with a nuclear time step of 0.5
fs and a local diabatization formalism\cite{PGPBPL19_LocalDiab} based on maximization of
the overlap matrix in two consecutive steps. 
The transition
probabilities between states were evaluated using the flux
formalism proposed by Mitric et al\cite{Mitric_FISH} in the adiabatic picture. 
An energy-based
decoherence correction was used with a parameter of 0.1
hartree\cite{GP07}.  
A set of 10000 initial
geometries and velocities were sampled considering the
Hessian of the MP2 optimization around the equilibrium
geometry of CB using a harmonic oscillator
Wigner distribution in Cartesian coordinates. 
Subsequently, we performed single point (SP) calculations at XMS-CAS(8,8)/aug-cc-pvdz level averaging over 4 singles and 4 triplets including SOCs to compute the energies and oscillator strengths for each initial condition to model the absorption spectrum. 

To mimic the wavepacket at time zero upon excitation, we narrow the initial conditions to the energy range matching the experimental pump pulse, i.e., 200 nm (6.2 eV) and $\sim 80$ fs cross-correlation. The excitation window corresponds to the $n\rightarrow3s$ transition. 
An ensemble of 200 trajectories falling in the bandwidth at $6.20\pm$0.01 eV) was selected for the dynamics calculations. The trajectories were propagated for $800$ fs starting from the bright spin-adiabatic (i.e. the eigenstate of the total Hamiltonian) excited state with dominant S$_2$ contribution. 
As observed in Table \ref{table:elec_struc}, the triplet states have energies similar to the singlet states. 
As a result, the numbering of the initial spin-adiabatic state varies across different trajectories, as the T$_1$ and T$_2$ triplets may lie either above or below the S$_2$ state depending on the geometry. Hence, our trajectories are starting from spin-adiabatic states $\vert8\rangle$ and $\vert11\rangle$.

\subsection{Non adiabatic dynamics analysis\label{sec:photoprods}}
We analyzed the trajectories to compute the time-dependent populations and characterize the different photoproducts formed during the dynamics after excitation. The time evolution of the state populations was
determined by averaging over all the trajectories in two different ways.

% The count of trajectories can only be done in the adiabatic states (spin-adiabatic)
First, we evaluate the spin-adiabatic populations as the average of active states, i.e., the fraction of trajectories in the ensemble that are evolving on a given state $k$ as $\Pi_k(t)=\sum_j^{N_{traj}}N_k^{(j)}(t)/N_{traj}$, where $k$ denotes the spin-adiabatic state, $j$ labels the nuclear trajectory and $N_{traj}$ is the total number of trajectories. 
% Here you are including a rho factor... the density can be calculated in all pictures. Since the previous one was the spin-adiabatic, probably you wat to speak here of the spin-diabatic
Alternatively, we average the electronic populations on each spin-diabatic state (i.e. states that are eigenvalues of the spin-free Hamiltonian) as $\rho_k(t)=\sum_{j}^{N_{traj}}\rho_k^{j}(t)/N_{traj}$, where $\rho_k^j(t)$ is the time-dependent electronic population on a given state $k$ for each nuclear trajectory $j$. 
%%??
%The latter equation was also utilized to evaluate the time-dependent populations of the spin-adiabatic states. 
We carefully monitored the total energy conservation in each trajectory by evaluating the total energy difference at each time step relative to the total energy at the initial time. We noticed that the total energy is well conserved until the CB opens through $\alpha$-cleavage and some trajectories become unstable, observing total energy jumps larger than $1$ eV.  

To compute the time-dependent populations and electron diffraction signals relative to the different photoproducts, we categorized the trajectories by visual analysis as: \textbf{C3} products (CO+C$_3$H$_6$), \textbf{C2} products (CH$_2$CO+C$_2$H$_2$), open structure and cyclobutanone. The \textbf{C3} category can further be subdivided into cyclopropanone + CO and propene + CO. For each category, we also computed the relative time-dependent populations.
%using the same procedure described earlier, where now $N_{traj}$ denotes the total number of trajectories in each category. 
%along with their corresponding time-dependent populations, are depicted in Fig. \ref{fig:photoproducts}. 
Furthermore, we determined the relative formation fraction for each photoproduct $i$ at the end of the propagation as $P_i(\%)=N_i/N_{traj}\times100$, where $N_i$ represents the number of trajectories resulting in a specific $i$ photoproduct at the end of the trajectory propagation. These results allow us to calculate the \textbf{C2}:\textbf{C3} branching ratio for comparison with previous theoretical and experimental findings.

\subsection{\label{sec:TSH}Ultrafast electron diffraction patterns}

From our trajectory surface hopping simulations, we calculate the theoretical time-resolved difference pair distribution function ($\Delta$PDF(t)). Each trajectory provides atomic positions as a function of delay, enabling computation of pair distribution functions (PDFs) for time-dependent molecular geometries. The 1D diffraction patterns are simulated using the independent atom model (IAM) for each molecular geometry and then converted to PDFs. Detailed methodology and equations can be found in Refs. \citenum{PDF_calc, PRX_Weinacht, UED_CHD_PDF}. The difference pair distribution function is obtained by subtracting the scattering signal at time zero providing a clearer observation of the different relaxation pathways. This signal is computed both for the ground state equilibrium geometry and the ensemble average of initial condition geometries. Comparing the two methods, we find insignificant differences in the PDFs (see Fig. \ref{fig:PDF_geom0}). Although the sampling effect has minimal impact on the PDFs, we utilize the signal from the average distribution in PDF computation.

\subsection{Details about the upcoming experiment}
A detailed description of  the experiment is available online\cite{experim}. Here we provide a short description with the main details of the time-resolved MeV-ultrafast electron diffraction (MeV-UED) measurements performed on cyclobutanone. 
Cyclobutanone will be irradiated with 200 nm light ($\approx$80 fs cross-
correlation) and electron diffraction images will be obtained with 150 fs time resolution (FWHM) and 0.6 Å spatial resolution $2\pi/S_{\mathrm{max}}$), covering a scattering vector S range from 1 to 10 Å$^{-1}$. 
The excitation is believed to target a Rydberg (3s) excited state (i.e., $n\rightarrow3s$) and not the $n\rightarrow\pi^*$ state (280 nm).  
The experimental setup will capture diffraction images over time delays ranging from -1 ps to several picoseconds, with variable step sizes. Specifically, the region around time zero (-200 fs to 200 fs) will be scanned with a step size of 30 fs, while longer positive delays will be scanned with step sizes up to several picoseconds. It is highly likely that one or more triplet electronic states will play a role in the dynamics.
\begin{figure}
\includegraphics[width=0.5\textwidth]{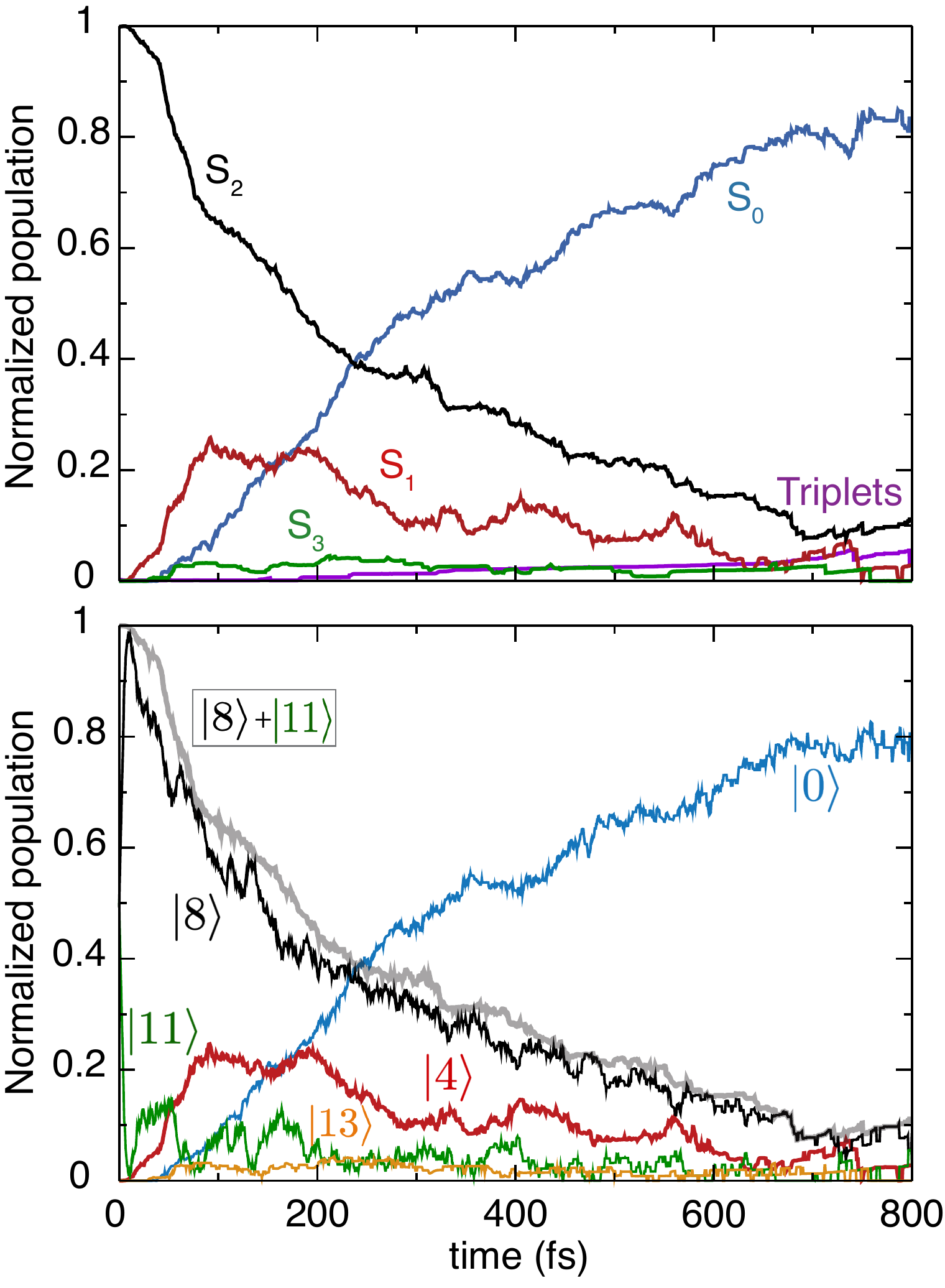}
\caption{Time evolution of the electronic populations in the spin-diabatic (spin-free) states (top) and the spin-adiabatic states (bottom) after excitation to the $n\rightarrow3s$ band calculated at XMS-CASPT2(8,8)/aug-cc-pvdz level. The thicker grey  line represents the sum of the population in the spin-adiabatic $\vert8\rangle$ and $\vert11\rangle$ states. It closely resembles the population trace of the spin-diabatic S$_2$ state, suggesting minimal mixing between states with different spin multiplicities. This observation is further supported by the negligible population within the triplet diabatic manifold.}
\label{fig:pop}
\end{figure}

\section{Results and discussion}\label{sec:results}
\subsection{Excited states of cyclobutanone: Absorption spectrum and validation of the electronic structure method}

% We will start our discussion by validating the electronic structure method selected for the non adiabatic dynamics and \textcolor{red}{...} calculations.  
 Experimentally, the band maximum for the $n\rightarrow\pi^*$ transition has been observed at $4.4$ eV\cite{hemminger_laser-excited_1973}, well separated from the $n\rightarrow3s$ transition, found at $6.29$ eV, while the origin is at $\sim6.11$ eV\cite{otoole_vacuum-ultraviolet_1991,drurylessard_ring_1978}. 
 This transition is the target of the experimental $200$ nm pump pulse.
 The $3p$ absorption band is very weak and has its origin on the transition to the lowest $3p$ Rydberg state appears at $6.94$ eV, with a maximum at $\sim7.25$ eV, although this remains somewhat uncertain due to the presence of a nearby, intense $3d$ absorption band\cite{otoole_vacuum-ultraviolet_1991}.
\begin{figure}
\includegraphics[width=0.8\textwidth]{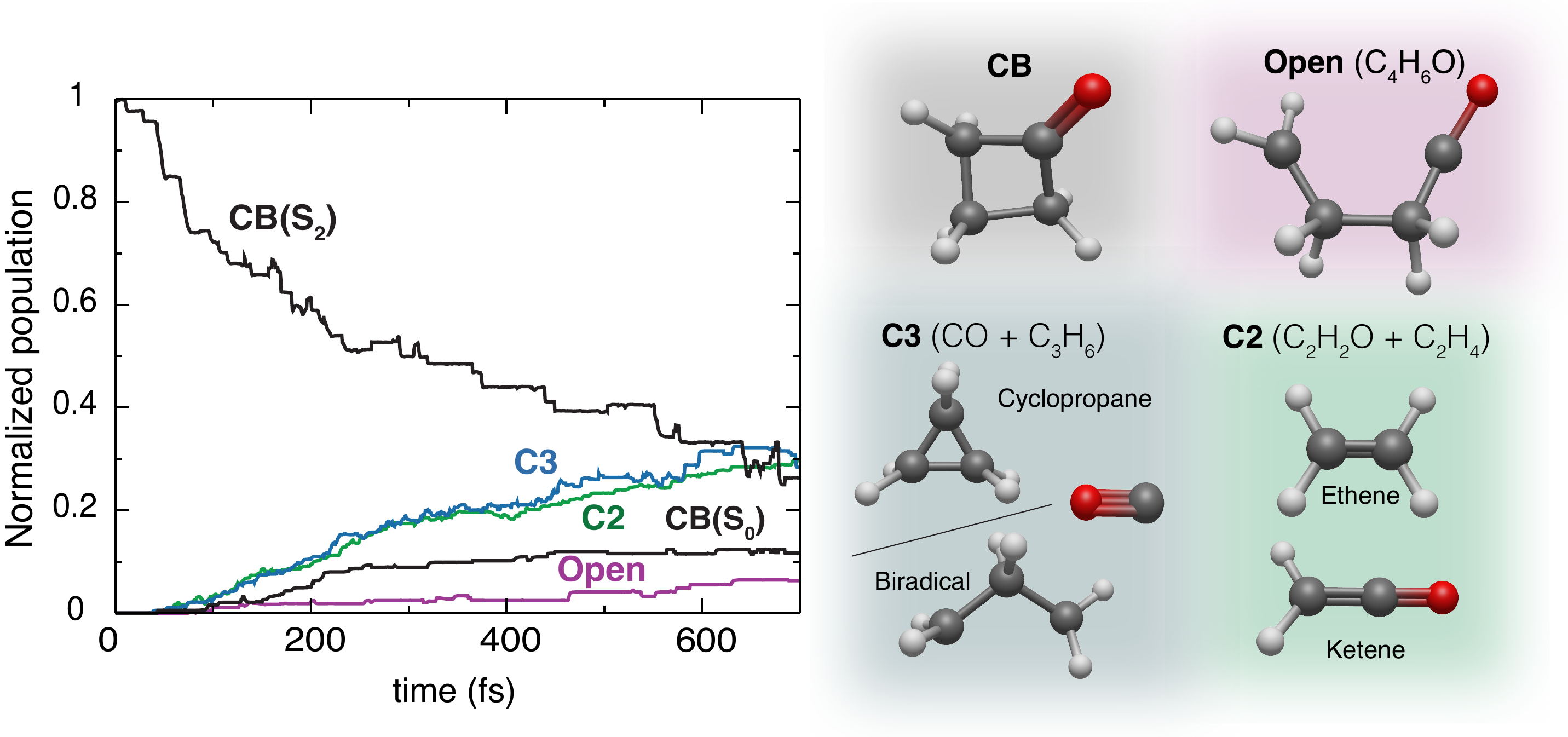}
\caption{Time-dependent  electronic populations (left) relative to the trajectories that exhibit the following structures: \textbf{C3} photoproducts (CO + C$_3$H$_6$) , \textbf{C2} photoproducts (C$_2$H$_2$O + C$_2$H$_4$), cyclobutanone, and the intermediate 4-carbon open structure in the first excited state. The population trace labeled as \textbf{CB(S$_2$)} refers to the average electronic population in the S$_2$ state of the trajectories conserving the ring structure at the end of the propagation. On the right, we display the structures of the four different categories into which we divide our photoproducts. The color code correlates the electronic populations with each category: \textbf{C2} (green), \textbf{C3} (blue), open structure (red), and cyclobutanone (black).}
\label{fig:photoproducts}
\end{figure}

Let us first start by assigning the character of the computed excited states and validate the selected electronic structure method. We have computed the first four vertical excitation energies at the equilibrium geometry of the ground electronic state of CB. 
Given that our active space is restricted to the $3s$ Rydberg orbital, transitions involving the $3p$ or $3d$ Rydberg orbitals cannot be described. Thus, the four singlet states computed correspond to transitions from the oxygen lone pair to the $\pi^*$ (S$_1$ state) and $3s$ Rydberg orbitals (S$_2$ state) ($n\rightarrow\pi^*$ and $n\rightarrow 3s$), as well as from the $\pi$ orbital to the $3s$ orbital (S$_3$). 
%%%%%
In the case of the triplet states, T$_1$, T$_3$ and T$_4$ have the same character than S$_1$, S$_2$ and S$_3$ whereas T$_2$ corresponds to the $\pi\rightarrow\pi^*$ excitation.
%%%%%%
We initially focus on the lowest excited states, namely S$_1$, S$_2$, T$_1$, T$_2$ and T$_3$.
Inspection of Table\ref{table:elec_struc} reveals that the transitions computed at SA(4/4)-CASSCF/aug-cc-pvdz level with (4,4) and (8,8) active spaces exhibit identical  energy values, showing the best agreement with experimental data, and differ by up to 0.35 eV  compared to SA-CAS(12,12) energies. A similar trend is observed in the XMS-CASPT2 energies, with a slightly more notable discrepancy in the $n\rightarrow 3s$ transition. Specifically, the S$_1$ XMS-CASPT2 energies are red-shifted relative to experimental values, contrasting with slightly higher values observed for S$_2$ XMS-CASPT2 energies in (4,4) active space. The S$_2$ SA-CASSCF(8,8) energies matches the experimental value.
Turning now to higher excited states corresponding to the $\pi\rightarrow 3s$ transition, S$_3$ and T$_3$, we note that the predicted SA-CASSCF(4,4) and  SA-CASSCF(8,8) energies are smaller than the SA-CAS(12,12) values, whereas the opposite behavior is observed in XMS-CASPT2 calculations. Moreover, these discrepancies across active spaces are larger ($\sim1$ eV) than those observed for the lower excited states. 

As detailed in Sec.\ref{sec:ElectronicStruc}, we also performed single point SA(4/4)-CASSCF(4,4) and XMS-CASPT2 calculations using the ano-rcc-pvdz basis set to account for relativistic effects. The energies deviate by approximately 0.7 eV from the experimental value. Given that this transition represents the target excited state of the pump pulse, we opted to exclude this basis set from our calculations. This decision was made because the initial state and subsequent dynamics would not be accurately described using this basis set.
\begin{figure}
\includegraphics[width=0.6\textwidth]{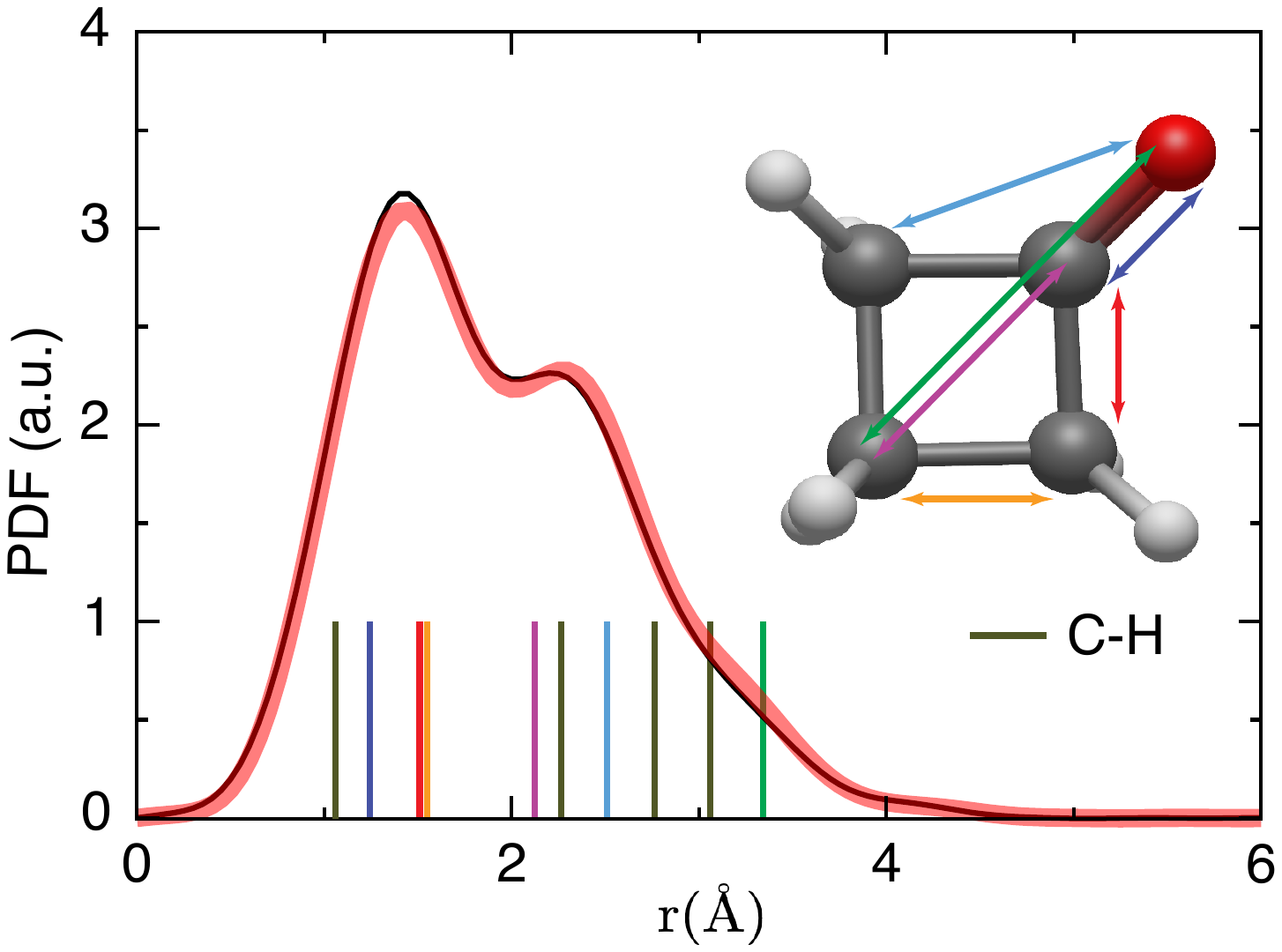}
\caption{Pair distribution function of cyclobutanone for the optimized structure (black) and the average structure from 200 initial conditions sampled from a Wigner distribution. The colored vertical lines indicate the atom-to-atom distances in the cyclobutanone ring and can be related to individual contributions to the PDF signal from different atomic pairs. The grey line represents the contribution from every C-H atomic pair. The colored arrows correlate the vertical lines with the atom-to-atom distances in CB.}
\label{fig:PDF_geom0}
\end{figure}

Despite the better agreement of CASSCF energies with the experimental values, we selected the XMS-CASPT2 method for our calculations, as dynamical correlation is essential to accurately describe bond breaking. As previously mentioned in Sec. \ref{sec:ElectronicStruc}, we selected an (8,8) active space for our calculations, striking a balance between accurately representing the excited states of interest and ensuring computational feasibility for our dynamics simulations. Additionally, we found that the aug-cc-pvdz and aug-cc-pvtz basis sets yield comparable vertical excitation energies, leading us to use the smaller basis set.
 
We now move to validate the chosen electronic structure method by simulating the absorption spectrum. To reproduce the experimentally observed spectrum,
excitation energies and oscillator strengths were computed for 10000 initial conditions at SA(4)-CASSCF/XMS-CASPT2/aug-cc-pvdz level of theory as detailed in Sec.\ref{sec:Methods}. Inspection of Fig. \ref{fig:abs_spectra} reveals 2 intense peaks centered at $6.2$ and at $9.5$, corresponding to the bright $n\rightarrow 3s$ and $\pi\rightarrow 3s$ transitions, and a weak absorption peak at $4.1$ eV, related to the $n\rightarrow \pi^*$ transition. The intensity of the peaks agrees with the computed oscillator strengths, with almost negligible values (below $10^{-5}$) for the $n\rightarrow\pi^*$ transition, and notably higher values for the brighter states. 
By comparing with the experimental spectrum, we confirm that the chosen method accurately reproduces the $n\rightarrow\pi^*$ and $n\rightarrow 3s$ electronic transitions, as the computed absorption spectrum closely matches the experimental one. Although the maximum of the theoretical absorption peaks appears slightly red-shifted compared to the experimental maximum values, the selected electronic structure method effectively captures the experimental peak width, particularly for the $n\rightarrow 3s$ transition.

\subsection{Non adiabatic dynamics}
We now proceed to analyze the nuclear trajectories and describe the relaxation dynamics upon excitation with a $200$ nm laser pulse.
Before delving into the discussion of the time-dependent population dynamics and different relaxation pathways, it is essential to address certain aspects concerning energy conservation and stability of the active space selected to perform the dynamics simulations. As described in Sec.\ref{sec:photoprods}, the total energy conservation is carefully monitored at individual trajectory level. Among the initial 200 trajectories, 30 trajectories exhibited a total energy variation exceeding $1$ eV, with 8 trajectories becoming unstable and crashing during the very early stages of propagation. The other 22 trajectories maintained total energy conservation until the $\alpha$-C-cleavage event. Following the ring opening, the energy of the 3s-Rydberg orbital increased, leading to its rotation with other orbitals and eventual removal from the active space. However, the energy jumps due to orbital rotations are observed when the trajectories are already evolving in the ground state. 
By the end of the propagation, we identified 192 non-crashing trajectories, with 137 still running at 350 fs, a point at which nearly $75\%$ of the transitions to the ground state have already occurred. Based on this, we can confidently assert that we have a sufficient number of trajectories to conduct a reliable statistical analysis.
We now turn to evaluate the time-dependent population dynamics starting from the photoexcited $3s$-Rydberg state and relaxation pathways leading to the formation of the different photoproducts.
Figure \ref{fig:pop} illustrates the average population dynamics of CB in both spin-diabatic (top) and spin-adiabatic states (bottom) as predicted by our trajectory surface hopping simulations throughout $800$ fs. The population traces depict the time-dependent average electronic population in each state. Upon comparing them with the time-dependent fraction of trajectories, we observe that the traces match, ensuring no violation of the internal consistency. The initial bright spin-adiabatic state, $\vert8\rangle$ or $\vert11\rangle$, corresponds mainly with the spin-diabaitc state S$_2$, with $>99\%$ population at $t=0$. 
This same characteristic holds for $\vert13\rangle$ and S$_3$, $\vert4\rangle$ and S$_1$, and $\vert 0\rangle$ and S$_0$ until about 200 fs, demonstrating minimal mixing between multiplets in the initial stages of the propagation. However, at around $200$ fs, we begin to observe a slight population transfer to the triplet states, reaching approximately 0.05 population by 800 fs. 
This could be considered negligible but can have an effect at longer propagation times. This observation suggests that triplet states should not be excluded in the dynamics to account for transitions at longer times. While the initial population transfer to the triplet states may appear minimal, its potential significance in longer propagation times, particularly when intersystem crossing (ISC) pathways become prominent, implies the necessity of considering triplet states in the dynamics to accurately capture the CB dynamics over extended time scales.

Examining the electronic populations of the spin-adiabatic states (lower panel), we notice a rapid transformation of the $\vert11\rangle$ state into the $\vert8\rangle$ state, with slight oscillations observed throughout the propagation window, especially during the first $200$ fs. Similarly to the impact in the spin-diabatic state ordering observed in the initial condition geometries, neighboring triplet states can also modify the order of spin-adiabatic states during the dynamics, depending on whether their energies are higher or lower than the state primarily dominated by S$_2$ contribution.
During the initial $75$ fs, we witness a rapid decay of the S$_2$ ($\vert8\rangle$+$\vert11\rangle$) state, accompanied by population transfer to S$_1$ ($\vert4\rangle$), reaching a maximum population of $0.25$. Subsequently, the population quickly funnels to the CB ground state S$_0$ ($\vert0\rangle$). At around $50$ fs, the ground state begins to populate, maintaining a steady increase until the end of the propagation. 
At $800$ fs, the population in the first excited state is practically zero, with $0.85$ population in the ground state. Minimal population transfer is observed to the S$_3$ state ($\vert13\rangle$); however, this scenario is considered unrealistic as, in practice, states corresponding to transitions from the oxygen $n$ orbital to 3p and 3d Rydberg orbitals would typically lie below the $\pi\rightarrow 3s$ state. 
% We cannot say that after highlight the importance of the triplets (with lower population)
%Nonetheless, given the negligible population, we believe this has minimal impact on the overall dynamics. 

To understand the possible mechanisms leading to different relaxation pathways and how the different photoproducts are formed, we analyzed the population dynamics by dividing our trajectories according to the molecular species formed at the end of our propagation. As mentioned in Sec.\ref{sec:photoprods}, we found the following photoproducts by visual analysis of the trajectories: \textbf{C3} products (CO+C$_3$H$_3$), \textbf{C2} products (C$_2$H$_2$O+C$_2$H$_3$), ciclobutanone (CB ring)  and an open structure formed after $\alpha$-C-cleavage. We observed two distinct time scales in the formation of the \textbf{C3} and \textbf{C2} products, which are directly influenced by the rate at which the open structure forms and population transfers to the ground state. In simpler terms, the speed at which the \textbf{C3} and \textbf{C2} products emerge depends on how quickly the open structure, which serves as the precursor to these products, develops. 
Once the ring opens, the open structure promptly transforms into the \textbf{C3} and \textbf{C2} products. This is further confirmed by the low fraction of trajectories found with an open structure at the end of the propagation. We will discuss  the fraction of formation of various photoproducts later in the text. 

But let us first continue examining the possible mechanisms of the different dissociation pathways.  
By inspecting the electronic populations at individual trajectory level, we noticed cases with almost direct population transfer to the ground electronic state from the S$_2$ state, suggesting the potential existence of a three-state conical intersection or that the S$_0$, S$_1$ and S$_2$ states are very close in energy for certain regions in configuration space. 
%A detailed analysis of the different relaxation pathways has been performed by J. Janoš \textit{et al.}\cite{janoš2024predicting} for the same prediction challenge. They found that the three electronic states considered become close in energy in the direct vicinity of the conical intersection of the open structure(S$_2$/S$_1$),  confirming our observations. 
We also observed a considerable number of trajectories maintaining the initial closed CB geometry, with some even reforming after the ring opening process. Referred to as "CB ring" trajectories, these trajectories exhibited a range of behaviors, with some remaining in the initial excited state while others transitioned to the ground state. Given our limited propagation time, we have not been able to thoroughly observe the evolution of these trajectories. However, based on the behavior of other trajectories, we expect that most of these trajectories will decay to the ground state, eventually leading to the $\alpha$-C-C cleavage and formation of \textbf{C3} and \textbf{C2} photoproducts.

These observations are supported by the relative populations of the photoproducts. 
The electronic population traces relative to each final product are shown in Fig. \ref{fig:photoproducts} with the structure of the respective photoproducts. We observe a gradual increase in the ground state populations of \textbf{C3} (blue line) and \textbf{C2} (green line) trajectories starting around 50 fs. However, a substantial ground state population leading to either "CB ring" or an open structure does not emerge until approximately 100 fs. This observation suggests that the  initial transfer of population from S$_2$ to S$_1$, and then to S$_0$, is primarily responsible for the rapid formation of \textbf{C3} and \textbf{C2}. Additionally, we notice that trajectories conserving the initial "CB ring" structure still retain some population in the initial excited state (CB(S$_2$) black line), but exhibit continuous decay, suggesting the eventual transfer of all the population to the ground state. Notably, some of the ground state population corresponds to a "CB ring" structure.
%In Ref. \cite{janoš2024predicting}, the authors found these hot trajectories to be highly unstable, eventually decomposing into one of the other products.

Finally, we summarize the ratio of the photoproducts after 800 fs propagation in Table\ref{table:frac_photop}. The low value of the relative population of the open structure confirms its role as an intermediate for the formation of the \textbf{C3} and \textbf{C2} photoproducts. The numbers also show a relatively large ratio of closed CB, due to to limitations of our 800 fs time propagation window. We expect this population to decrease at longer times. From the \textbf{C3} and \textbf{C2} distributions, we can also compute the \textbf{C3:C2} ratio. Our 55:45 ratio agrees remarkably well with the experimental 57:43 ratio observed by Trentelman {\it et al.}\cite{Trentelman_90} after photoexcitation with a 193 nm laser.

\begin{table}[]
%\resizebox{\columnwidth}{!}{%
\begin{tabular}{cc}
\hline\hline
\textbf{Product} & \textbf{Distribution (\%)} \\ \hline\hline
C2          & 31.2\%                     \\
C3               & 38.7\%                     \\
CB ring              & 22.6\%                     \\
CB open          & 7.5\%                     
\end{tabular}%
	\caption{Relative distribution of the photoproducts of cyclobutanone following irradiation at 200 nm as predicted after 800 fs propagation from our TSH/XMS-CASPT2 simulations}
\label{table:frac_photop}
\end{table}

\subsection{Ultrafast time-resolved electron diffraction}
We conclude our discussion by examining the electron diffraction patterns predicted from our dynamics simulations. We first start by examining the theoretical static pair-distribution function, PDF(r)  of ciclobutanone prior photoexcitation. In Fig. \ref{fig:PDF_geom0}, we depict the expected PDF for the initial geometry as the signal computed from the ground state equilibrium geometry (black) and the average signal originating from our 200  initial condition geometries used in our dynamics calculations. 

The PDF reveals two prominent peaks: one centered at 1.4 \text{\r{A}} and another less intense peak at 2.2 \text{\r{A}}.
Further analysis involves decomposing the PDF obtained  into contributions from different atom pairs within the molecule. We compare the simulated signal with the atom-to-atom distances at the optimized geometries, shown as vertical colored lines and as arrows in the CB molecule, in Fig. \ref{fig:PDF_geom0}. The initial peak at 1.4 \text{\r{A}} primarily results from bonded pairs of atoms, i.e., C-O, $\alpha$-C-C, $\beta$-C-C, and C-H bonds. The second peak at 2.2 \text{\r{A}} arises from non-bonded atom pairs, predominantly involving distances between the C or O of the carbonyl group and the opposite carbons in the cyclobutanone ring, as well as non-bonded C-H distances. The shoulder-like structure around 3.4 \text{\r{A}} originates from the distance between the oxygen and the opposite $\beta$-C.

\begin{figure}
\includegraphics[width=0.8\textwidth]{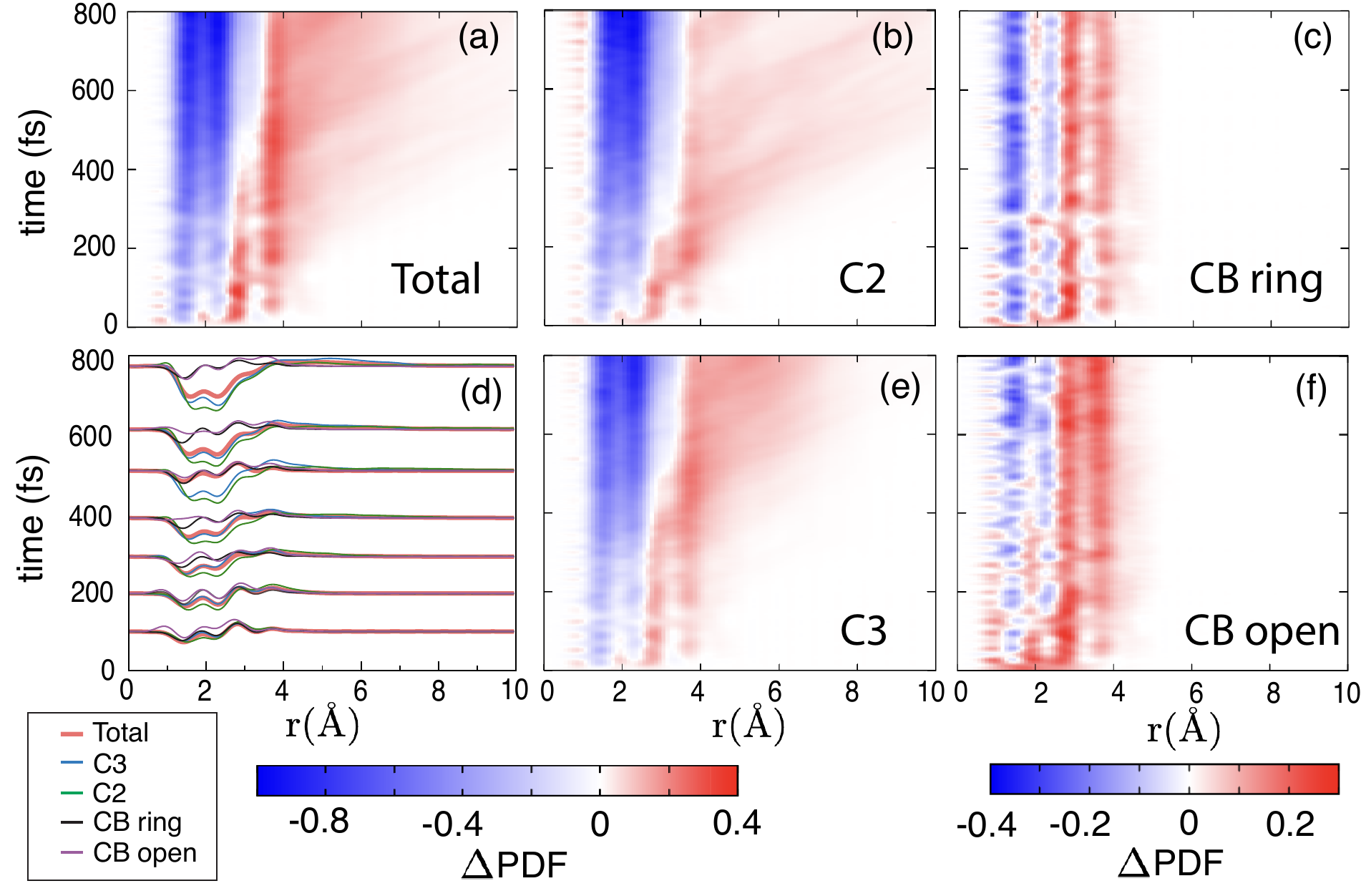}
\caption{Simulated time-resolved $\Delta$PDF(r,t) signals obtained from the swarm of TSH/XMS-CASPT2 trajectories. (a)Total signal averaged over all the trajectories. Signal  averaged over the trajectories following the (b) \textbf{C2}, (c) \textbf{C3} pathways and trajectories with (e) CB ring and (f) CB open structures at the end of the propagation. (d) Time slices of $\Delta$PDF(r,t) from 0 to 800 fs. Total signal (red), \textbf{C3} (blue), \textbf{C2}(green), CB ring (black) and CB open structure (purple). Note that the $\Delta$PDF(r,t) signals are not normalized with respect to the total trajectory number, just averaged over the number of trajectories in each category.  }
\label{Fig:TD_PDFs}
\end{figure}

We now look at the simulated time-resolved $\Delta$PDF(r,t) computed from our trajectory 
surface hopping calculations. 
Fig. \ref{Fig:TD_PDFs} shows the time variation of the $\Delta$PDF(r) for the total swarm of trajectories (a,d) and for the trajectories in each category of photoproducts (b,c,e,f). 

For all the $\Delta$PDF signals we can distinguish between positive ($\sim$2.5 and 8 \text{\r{A}} 
) and negative (between 1 and 2.5 \text{\r{A}}) features. Let us first look at the total $\Delta$PDF signal. During the first 50 fs we observe a very weak signal when the cyclobutanone relaxes from the S$_2$ to the S$_1$ state, common to all the $\Delta$PDFs for all the contributions. After the initial population transfer, a positive signal rises at 3.7-4 \text{\r{A}}, attributed to the ring-opening process and formation of the open structure. Fig. \ref{Fig:TD_PDFs} (f), the same signal is observed trajectories that maintain the open structure during the propagation, becoming more prominent at the end of the dynamics, confirming our hypothesis. At later times, from about 200 fs, the evolution of the positive signals towards larger $r$ values reflect the dissociation into \textbf{C3} and \textbf{C2} products. In addition, we observe a depletion of the feature around 2.5-3 \text{\r{A}} at about 500fs, followed by the appearance of a negative weak signal at the same pair distance, attributed to the formation  of the \textbf{C2} and \textbf{C3} photoproducts.

Looking at the categorized $\Delta$PDFs depicted in panels (b), (c), (e), and (f) of Fig. \ref{Fig:TD_PDFs}, we observe a similar evolution of the positive signal towards larger atomic pair distances in the \textbf{C3} and \textbf{C2} channels, absent in the $\Delta$PDFs related to CB ring and CB open structures.  Moreover, the presence of two prominent negative features between 1-2.5 \text{\r{A}} correlates with the bond breaking during dissociation. This bond breaking can be attributed to the dissociation into \textbf{C2} and \textbf{C3} photoproducts or the open structure. While this signal increases in intensity over time and is also evident in the \textbf{C2} and \textbf{C3} $\Delta$PDF plots, it remains relatively weak in the cases of CB open and CB ring structures. The $\Delta$PDF time-slices presented in Fig. \ref{Fig:TD_PDFs}(d) provide a clear visualization of the increasing intensity (in absolute value) of the two negative features between 1-2.5 \text{\r{A}}. From approximately 600 fs onwards, the overall $\Delta$PDF signal mirrors the shape of the \textbf{C2} and \textbf{C3} $\Delta$PDFs, with the intensity of these negative signals notably amplifying as the dynamics progress. Moreover, a negative shoulder emerges in the total $\Delta$PDFs at around 3 \text{\r{A}}, akin to the behavior observed in the \textbf{C2} and \textbf{C3} $\Delta$PDFs. The features observed in the time-resolved $\Delta$PDF are consistent with the population dynamics shown in Figs. \ref{fig:pop} and \ref{fig:photoproducts}. Following the initial transfer from S$_2$ to S$_1$, there is an increase in ground state population at approximately 50 fs, coinciding with the emergence of the \textbf{C2} and \textbf{C3} products and the appearance of two negative signals. The subsequent rise in positive signals at 3-4 \r{A} corresponds with the decrease in S$_2$ and S$_1$ populations, indicative of cyclobutanone dissociation into different photoproducts.
We can conclude that while the positive signal at large atomic pair distances indicative of dissociation may be subtle, posing challenges for direct observation on longer timescales, the highly intense feature at short pair distances will be distinctly identifiable in the UED signal, even at later times. To differentiate between the \textbf{C3}/\textbf{C2} dissociation channels and ring-opening, it will be crucial to capture the signal at short atomic pair distances (1-2.5 \r{A}), corresponding to a momentum space of $s=0-1$ \r{A}$^{-1}$.

\section{Conclusions}\label{sec:conc}
In this work, we present the results of trajectory surface hopping with high-level XMS-CASPT2 simulations to investigate the dynamics of cyclobutanone following photoexcitation with a 200 nm pulse. We then predict the time-resolved MeV-UED signals, which will be compared to data from an upcoming experiment. 

The relaxation dynamics of cyclobutanone upon excitation to the 3s-Rydberg state progress through  S$_2\rightarrow$S$_1\rightarrow$S$_0$, leading to the formation of various photoproducts. At the 800 fs, approximately 80\% of the population resides in the ground state, with 10\% remaining in the initial S$_2$ excited state. Notably, some population transfer to triplet states is observed, suggesting the  potential implications and importance in the  dynamics at longer timescales and photoproduct formation.
Several photoproducts have been identified, with CO and C$_3$H$_3$ (in the form of cyclopropanone or a 3-carbon birradical) being the predominant species (38.7\%). This is closely followed by ethene and ketene (31.2\%), resulting from $\alpha$-C-cleavage and ring-opening of cyclobutanone, leading to a \textbf{C3}:\textbf{C2} ratio of 55:45.
The formation of these photoproducts initiates early in the dynamics (at 50 fs) and continues throughout the propagation. Approximately 22.6\% of cyclobutanone remains either in the ground state or in the initial excited state by the end of the propagation period.

The primary aim of this investigation is to anticipate the MeV-UED patterns associated with the photochemistry of cyclobutanone following excitation at 200 nm. We have calculated the time-resolved difference pair distribution functions of cyclobutatone from our swarm of trajectories. Additionally, we provide analogous electron diffraction signals for the diverse observed photoproducts.  Our results suggest that MeV-UED measurements will yield electron diffraction signals with characteristic features associated to the photochemistry of cyclobutanone, facilitating the experimental monitoring of ring opening dynamics and the generation of \textbf{C3} and \textbf{C2} photoproducts. 

The time constraints inherent in the prediction challenge posed several hurdles for accurately predicting the experimental outcomes. Given that the proposed experiment involves excitation to a 3s-Rydberg state and potential dissociation processes, along with the possible involvement of triplet states in the dynamics, an accurate description required the use of large basis sets with diffuse functions, multireference methods, and the proper inclusion of dynamical correlation and spin-orbit couplings. 
To address these challenges, we employed the XMS-CASPT2 method, which led to highly computationally-demanding simulations.
This method is highly sensitive to the precise selection of its underlying active space and the number of electronic states considered. To balance accuracy and efficiency, we limited our active space to the 3s-Rydberg orbital, excluding higher states that were energetically close. Unfortunately, this limitation resulted in some instabilities in the active space, leading to the breakdown of certain trajectories.
Furthermore, due to time constraints and the computational demands of the XMS-CASPT2 method, we were unable to propagate the trajectories to longer time scales. As a result, the accuracy of the discussed results, particularly in terms of the formation ratio of the different photoproducts, may not be completely accurate.

Despite these limitations, our study demonstrates the utility of on-the-fly trajectory surface hopping simulations combined with high-level \textit{ab initio} methods for providing a clear and direct comparison with experimental observables. For this reason, these methods stand out  as a powerful tool for predicting and understanding phenomena observed in experiments.

\begin{acknowledgments}
JGV thanks the PID2022-138288NB-C32 and PID2019-106732GB-I00 projects funded by MCIN/AEI/10.13039/ 501100011033 and the European Union “NextGenerationEU”/PRTRMICINN programs. PVZ thanks the National Science Foundation for its support under the Grants CHE-2054616 and CHE-2054604. We acknowledge the Simons Foundation for the computational resources  that were used in this research. 
\end{acknowledgments}

\section*{Data Availability Statement}
The data that support the findings of this study are available from the corresponding author upon request. 
The temporary arXiV submission
identifier is: submit/5412196.
\section*{Author declarations}
\section*{Conflict of interest}

The authors have no conflicts to disclose.

\section*{ Author Contributions}
{\bf Patricia Vindel-Zandbergen}: Conceptualization (lead); Formal analysis (lead); Visualization (lead); Software (supporting); Investigation (lead); Writing original draft (lead); Writing – review \& editing (lead). {\bf Jesús Gonzalez-Vázquez}: Software (lead); Writing original draft (supporting); Formal analysis (supporting)
\bibliographystyle{jcp}
\bibliography{ref_na_2, CB,theory,ref, ref_na}

\newpage
\section{Supplementary Material}\label{sec:SI}
\beginsupplement

\subsection{Active space and molecular orbitals}

\begin{figure}[h!]
\includegraphics[width=0.8\textwidth]{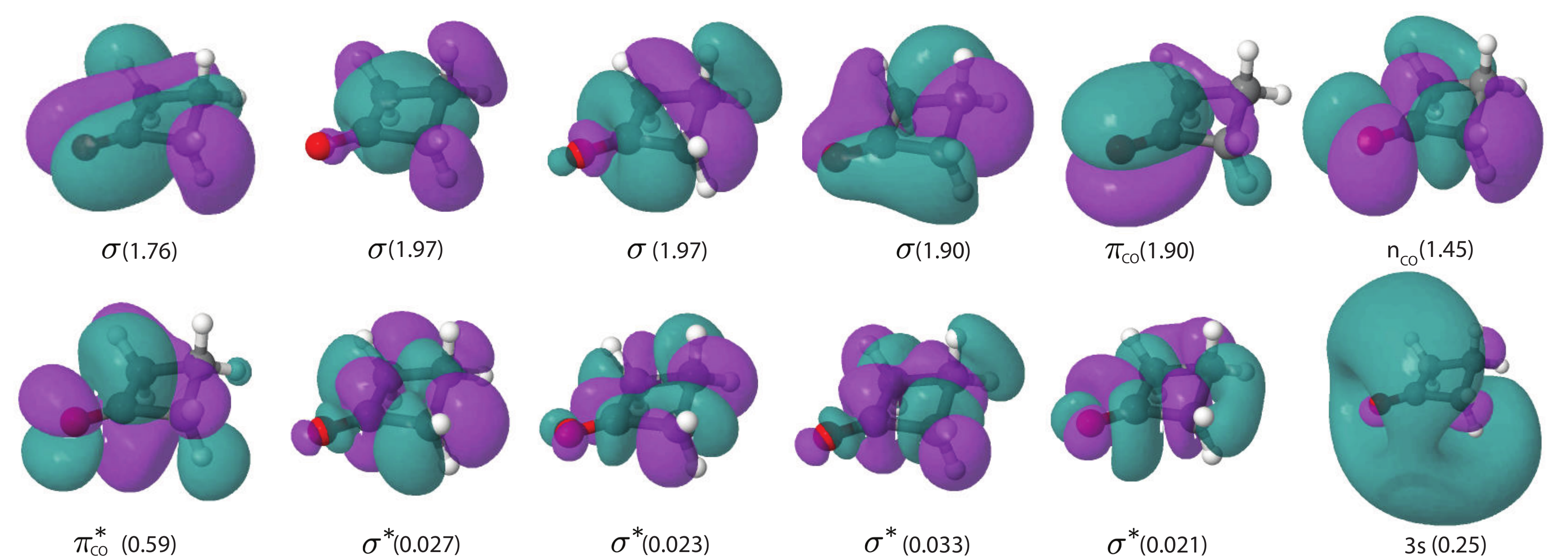}
\caption{Molecular orbitals of cyclobutanone included in the active space in the  SA(4)-CASSCF(12,12) and XMS-CASPT2 calculations. The character of each orbital is detailed and the occupation numbers are given within parenthesis}
\label{Fig:orbitals_CB_12}
\end{figure}

\begin{figure}[h!]
\includegraphics[width=0.6\textwidth]{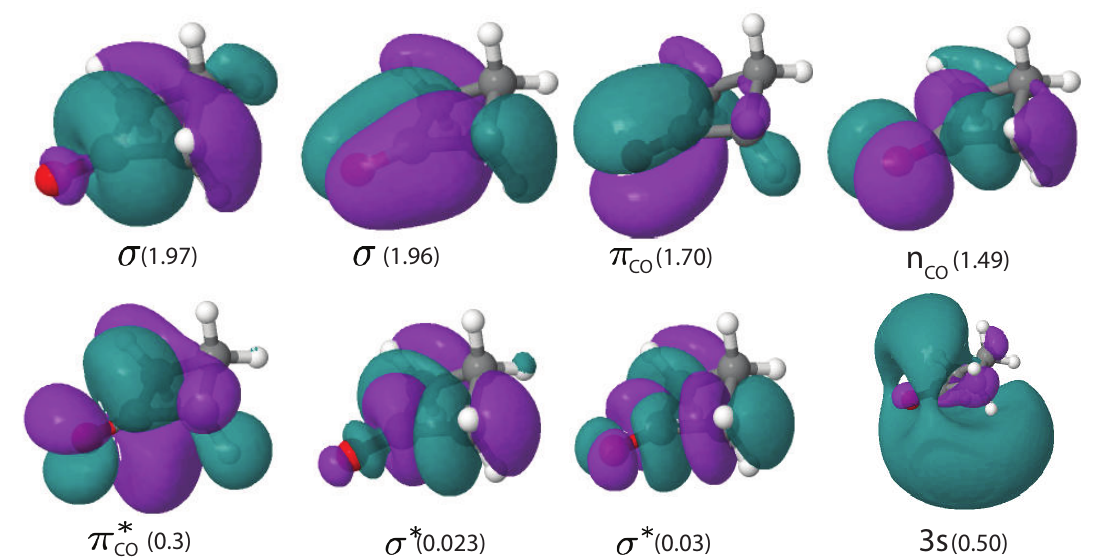}
\caption{Molecular orbitals of cyclobutanone included in the active space in the SA(4)-CASSCF(8,8) and XMS-CASPT2 calculations. For each orbital, the character is detailed and the occupation numbers are given within parenthesis. The active space in the SA(4)-CASSCF(4,4) includes the $\pi,\pi*,n$ and 3s orbitals with the same occupation numbers.}
\label{Fig:orbitals_CB_8}
\end{figure}

\end{document}